\documentclass[fleqn,usenatbib]{mnras}

\usepackage{newtxtext,newtxmath}

\usepackage{xcolor}

\usepackage[T1]{fontenc}
\usepackage{ae,aecompl}

\usepackage{graphicx}	
\usepackage{amsmath}	
\usepackage{bm}
\usepackage{verbatim}   
\usepackage[normalem]{ulem} 
\usepackage{units}

\newcommand{\cqg}{Class. Quant. Grav.}

\title[$(q,\chi_{\rm eff})$ anti-correlation in  AGN]{
LIGO--Virgo correlations between mass ratio and effective inspiral spin: testing the active galactic nuclei channel}

\author[B.McKernan et al.]{B. McKernan$^{1,2,3,4}$\thanks{E-mail:bmckernan at amnh.org (BMcK)}, K.E.S. Ford$^{1,2,3,4}$, T. Callister$^{2}$, W.M. Farr$^{2,5}$, R.O'Shaughnessy$^{6}$, R.Smith$^{7,8}$, 
\newauthor E. Thrane$^{7,8}$ \& A.Vajpeyi$^{7,8}$ \\
$^{1}$Department of Astrophysics, American Museum of Natural History, New York, NY 10024, USA\\
$^{2}$Center for Computational Astrophysics, Flatiron Institute, New York, NY 10010, USA\\
$^{3}$Graduate Center, City University of New York, 365 5th Avenue, New York, NY 10016, USA\\
$^{4}$Department of Science, BMCC, City University of New York, New York, NY 10007, USA\\
$^{5}$Department of Physics and Astronomy, Stony Brook University, Stony Brook NY 11794, USA\\
$^{6}$ Center for Computational Relativity and Gravitation, Rochester Institute of Technology, Rochester, NY 14623, USA\\
$^{7}$School of Physics and Astronomy, Monash University, Clayton, VIC 3800, Australia\\
$^{8}$ OzGrav: The ARC Centre of Excellence for Gravitational Wave Discovery, Clayton, VIC 3800, Australia\\
}

\date{Accepted XXX. Received YYY; in original form ZZZ}

\pubyear{2020}

\begin{document}
\label{firstpage}
\pagerange{\pageref{firstpage}--\pageref{lastpage}}
\maketitle

\begin{abstract}

Observations by LIGO--Virgo of binary black hole mergers suggest a possible anti-correlation between black hole mass ratio ($q=m_{2}/m_{1}$) and the effective inspiral spin parameter $\chi_{\rm eff}$, the mass-weighted spin projection onto the binary orbital angular momentum \citep{Tom2021}.
We show that such an anti-correlation can naturally occur for binary black holes assembled in active galactic nuclei (AGN) due to spherical and planar symmetry-breaking effects.
We describe a phenomenological model in which: 1) heavier black holes live in the AGN disk and tend to spin up into alignment with the disk; 2) lighter black holes with random spin orientations live in the nuclear spheroid; 3) the AGN disk is dense enough to rapidly capture a fraction of the spheroid component. but small in radial extent to limit the number of bulk disk mergers; 4) migration within the disk is non-uniform, likely disrupted by feedback from migrators or disk turbulence; 5) dynamical encounters in the disk are common and preferentially disrupt binaries that are retrograde around their center of mass, particularly at stalling orbits, or traps.
This model may explain trends in LIGO--Virgo data while offering falsifiable predictions. Comparisons of predictions in ($q,\chi_{\rm eff}$) parameter space for the different channels may allow us to distinguish their fractional contributions to the observed merger rates.
\end{abstract}


\section{Introduction}
Active galactic nuclei (AGN) are powered by the accretion of gas onto a supermassive black hole (SMBH). A dense population of stars and stellar remnants is also expected to orbit the SMBH  \citep{1993ApJ...408..496M,2000ApJ...545..847M,2018Natur.556...70H,2018MNRAS.478.4030G}. A fraction of  this orbiting spheroid population is captured by the AGN over the disk lifetime \citep{Fabj20,MacLeodLin20}, leading to a large population of embedded orbiters within AGN disks. The embedded population experiences gas torques which leads to migration and close dynamical encounters. If binaries form as a result of these encounters, they can be hardened or softened by gas torques or further dynamical interactions \citep[e.g.][]{McK12,Leigh18}. Hardening of  binaries to merger yields gravitational wave (GW) signals  detectable with LIGO--Virgo \citep[e.g.][]{McK14,Bartos17,Stone17}. Since AGN live in deep potential wells, merger products are easily retained in spite of $\sim\unit[1000]{km\, s^{-1}}$ GW kicks \citep{Gerosa19}, yielding a mass hierarchy due to repeated mergers,
possibly over multiple AGN episodes.

The properties of AGN are poorly constrained by electromagnetic (EM)
observations, with multiple orders of magnitude variation in density and disk
aspect ratio allowed within common models \citep{Sirko03,TQM05} and orders of
magnitude uncertainty in AGN lifetimes \citep{Schawinski15}. This necessarily
yields a very large range of predicted compact binary merger rates from this
channel \citep{McK18,Grobner20}. Conversely, GW observations of mergers from
this channel allow us to `reverse engineer' properties of both the AGN disk and
the nuclear star cluster \citep[e.g.][]{McK18}.

Detailed simulations of merging populations in the AGN channel include $N$-body simulations \citep[e.g.][]{Secunda19,Secunda20a}, Monte Carlo simulations \citep[e.g.][]{Yang19,Tagawa19,McK20a,Tagawa20b,McK20b,Yang20,Tagawa21} and hydrodynamic simulations \citep[e.g.][]{Li21,Derdzinski21}. Broadly, BBH systems can merge in the bulk AGN disk, outside the disk if ejected by dynamical encounters,
and at special locations in the AGN disk such as a migration trap \citep{Bellovary16}, where most migrators arrive from larger disk radii \citep{Secunda20b,McK20a}. The predicted population properties of the AGN channel, including mass and spin distributions as well as rates are broadly consistent with the results of O3a \citep{o3a_pop}, including the relative rates of highly asymmetric mass ratio events such as GW190814 \citep{GW190814}, or intermediate mass black hole formation events such as GW190521 \citep{GW190521,GW190521_implications}. Uniquely, EM counterparts to BBH mergers may occur in this channel \citep{McK19a,Graham20a}, although confirming candidates is difficult \citep{Ashton2020,Palmese21}.
However, distinguishing between different merger channels is difficult based on mass and spin distributions alone.

Here we discuss the phenomenological implications for the AGN channel from the recent claim that there is a negative correlation between the mass ratio ($q\equiv m_2/m_1$) and the effective inspiral spin parameter $\chi_{\rm eff}$ found in the LIGO--Virgo O3a data \citep{Tom2021}. 

\section{An anti-correlation between mass ratio and effective inspiral spin parameter in BBH mergers}

With the conclusion of their O3a observing run, the Advanced LIGO~\citep{ligo} and Virgo~\citep{virgo} experiments have observed gravitational waves from 48 candidate BBH detections~\citep{gwtc2}, with additional candidates reported by independent re-analyses of LIGO--Virgo data~\citep{2020PhRvD.101h3030V,ogc3}.
Using hierarchical Bayesian inference, \cite{Tom2021} explored the degree of correlation between BBH mass ratio $q \equiv m_2/m_1$ and the effective inspiral spin parameter
    \begin{equation}
    \chi_\mathrm{eff} = \frac{ \vec{\chi}_1 + q \vec{\chi}_2 }{1+q}\cdot \hat{L}_{\rm b},
    \end{equation}
defined as the mass-weighted average of a binary's dimensionless component spins $\vec{\chi}_i$, projected onto the direction of the binary's orbital angular momentum $\hat{L}_{\rm b}$.
Modeling the $\chi_\mathrm{eff}$ distribution at fixed $q$ as a truncated Gaussian with peak value $\mu$ and width $\sigma$, \cite{Tom2021} measured the slopes $d\mu/dq$ and $d\log \sigma/dq$ with which the distribution's mean and (log) standard deviation vary with $q$.
They found $d\mu/dq = -0.46^{+0.29}_{-0.28}$ (median and 90\% credible uncertainties) and constrained $d\mu/dq<0$ at 99\% credibility.
The data does not determine whether the \textit{width} of the $\chi_\mathrm{eff}$ distribution varies with BBH mass ratio, and remain consistent with a $\chi_\mathrm{eff}$ distribution of width $\sigma \approx 0.1$ across all mass ratios.

The strong preference for negative $d\mu/dq$ may indicate that BBH mass ratios and effective spins are anti-correlated, with unequal-mass events possessing (on average) positive $\chi_\mathrm{eff}$ and equal-mass binaries possessing small or vanishing $\chi_\mathrm{eff}$.
The anticipated detection of several tens of additional BBH detections in LIGO \& Virgo's O3b observing run, however, will further clarify the relationship between $q$ and $\chi_\mathrm{eff}$.

\section{Phenomenological Considerations}
Here we shall make the assumption that all BBH mergers come from AGN. We can then determine what regions of parameter space---and what physical conditions---we require to produce some kind of anti-correlation between ($q,\chi_{\rm eff}$). In the future, a detailed parameter space study could determine the permitted values for AGN channel parameters (and their degeneracies), as well as the degeneracies with various possible mixing fractions for other channels; however that is beyond the scope of this work.
In particular, we should bear in mind that the anti-correlation might involve only a fractional contribution from the AGN channel.
For example, AGN might dominate the production of asymmetric mass mergers involving higher masses and spin but might be sub-dominant at lower $\chi_{\rm eff}$ or $q \sim 1$ \citep{Gayathri21}.
In this case, it may be the \textit{mixture} between the AGN and other binary formation channels that gives rise to the observed correlation between mass ratio and spin.

\begin{figure}
\begin{center}
\includegraphics[width=0.9\linewidth,angle=0]{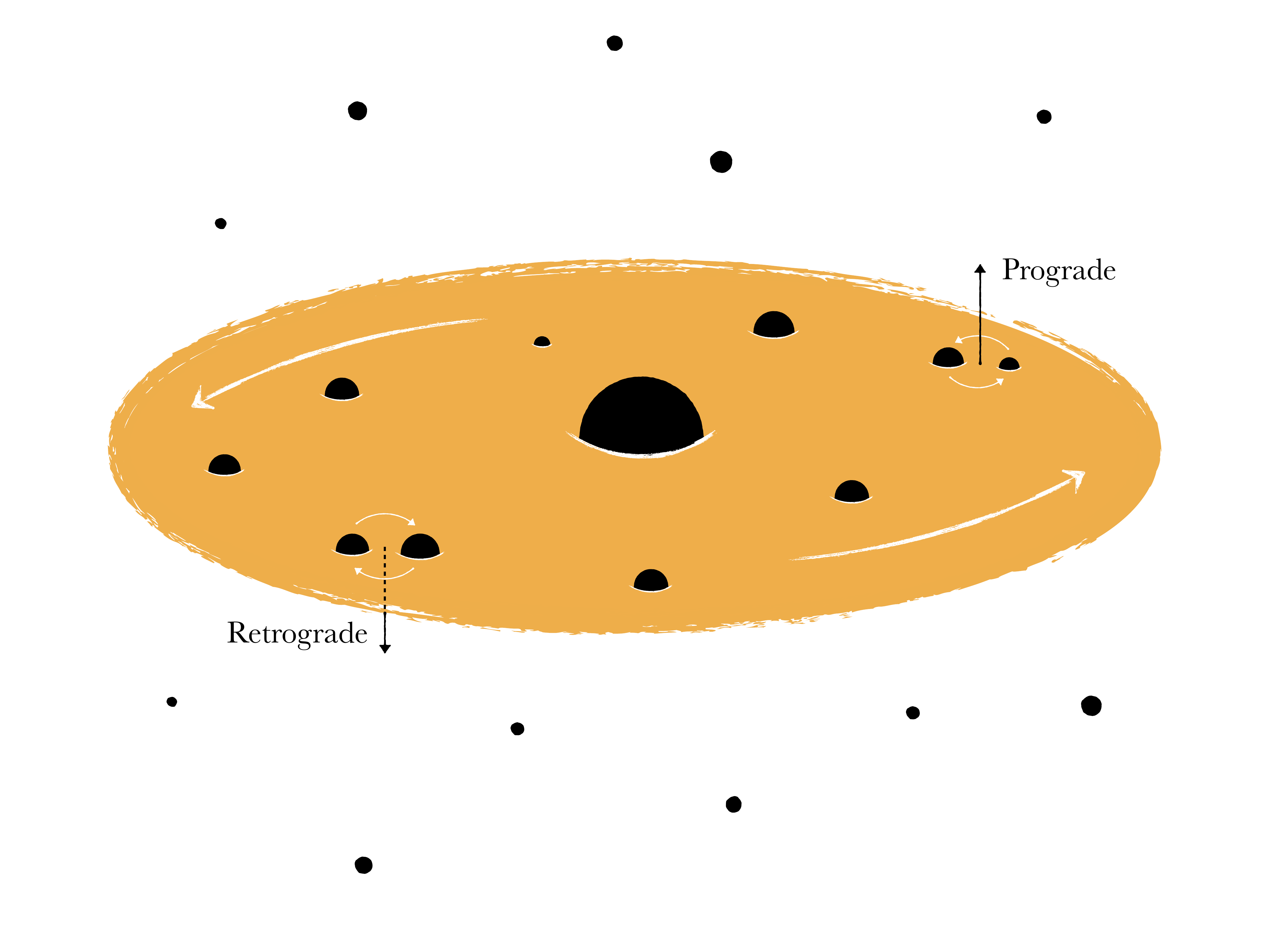}
\end{center}
\caption[Retrodiag]{Prograde versus Retrograde Binaries: Single black holes in an AGN disk that are orbiting the SMBH with their angular momentum aligned to the disk angular momentum $\vec{L}_{\rm d}$ will form binaries whose center of mass will continue to orbit with the disk gas. However, the orbital angular momentum of the binary around its own center of mass may be either prograde (aligned with $\vec{L}_{\rm d}$) \textit{or} retrograde (anti-aligned with $\vec{L}_{\rm d}$). We use the terms `prograde' and `retrograde' binary, respectively, for these two arrangements. Because single BH orbiting retrograde with respect to the AGN disk gas should have lifetimes short compared to the gas disk \citep{Secunda20b}, we do not expect binaries whose center of mass orbits in a retrograde sense with respect to the disk gas, and we neglect such objects entirely in our analysis. 
\label{fig:retrodiagram}}
\end{figure}

Below, we discuss properties of binaries and their encounters. We define `retrograde' (`prograde') binaries as those which rotate around their center of mass such that $\vec{L}_{\rm b} \cdot \vec{L}_\mathrm{disk} <0$ ($\vec{L}_{\rm b} \cdot L_\mathrm{disk} >0$)---see Figure \ref{fig:retrodiagram}. The hard-soft boundary for a binary is when the binary orbital energy is approximately equal to the average energy ($M_{3}\Delta v^{2}$) of encounters. Softening (hardening) encounters tend to increase (decrease) binary separation. A binary is `ionized' when $KE + PE > 0$ after an encounter.

\subsection{Symmetry breaking in AGN}
Classical dynamical merger channels (e.g. globular clusters and nuclear star clusters) generally assume  spherical symmetry. AGN are distinctive in that there are several sources of spherical symmetry breaking, some of which may play a role in the possible anti-correlation discussed here. We outline some of the basic concepts here and elaborate further below.

The AGN disk tends to break spherical mass symmetry in a galactic nucleus since 1) more massive BH are preferentially captured by the disk \citep{Fabj20}, 2) star formation in AGN disks should come with a top-heavy IMF \citep[e.g.][]{Yuri03} and 3) a combination of mass segregation/star formation in galactic nuclei may yield disky assemblies of massive BH \citep[e.g.][]{Alexander07}.

The symmetry of dynamical encounters in AGN (between a binary and a singleton) may be broken because: 1) spherical symmetry for encounters in the galactic nucleus is broken by the preference for $\vec{L}_{\rm b}$ to form parallel or anti-parallel to $\vec{L}_{\rm d}$, and disk capture of singleton orbiters then biases encounters towards co-planar arrangements 2) there may be regions in the inner disk where migration stalls, e.g.  migration traps, so encounters will typically arrive from the outer disk, breaking even the planar symmetry of encounters \citep{Bellovary16,Secunda20a}, 3) BH orbiting the SMBH on retrograde orbits may rapidly decay onto the SMBH, biasing orbits to prograde \citep{Secunda20b}, 4) Retrograde binaries have a different hard-soft boundary than prograde binaries for encounters that are prograde, 5) massive objects migrate faster in the disk so there is a mass bias in `catch up' encounters, 6) gas torques may preferentially harden/soften retrograde/prograde binaries, depending on the details of gas flow around the BBH \citep{Baruteau11,Li21}.

\subsection{Phenomenology}
\begin{figure*}
\begin{center}
\includegraphics[width=\linewidth,angle=0]{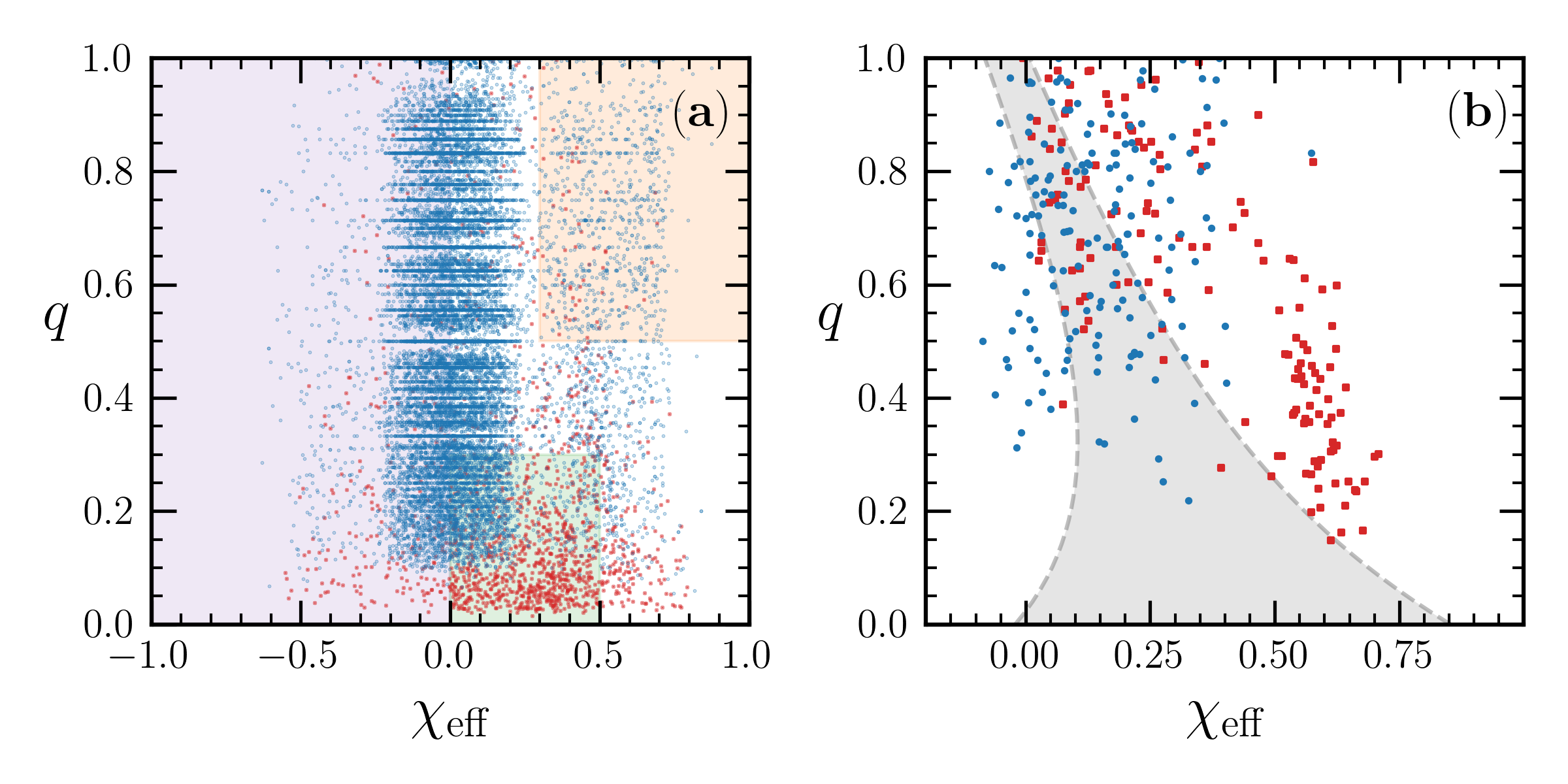}
\end{center}
\caption[Mergers]{
\textit{Left panel:} 
Mass ratio ($q=M_{2}/M_{1}$) as a function of $\chi_{\rm eff}$ for binary black hole mergers in the bulk disk from 100 realizations of the R7 model of \citet{McK20a}. Blue points correspond to mergers in the bulk disk, red points correspond to mergers at a trap. Horizontal structures correspond to artefacts of mass bins among early mergers (e.g. mergers between $5.0M_{\odot}+10.0M_{\odot}$ BH before any significant accretion has occurred). The results show no interesting correlations. In order to produce the $q-\chi_{\rm eff}$ anti-correlation observed in \citet{Tom2021}, there must be physical mechanisms for suppressing 3 types of mergers, highlighted in the plot and discussed in the text: 1) negative $\chi_{\rm eff}$ mergers (purple region); 2) mergers with $q
\ga 0.5$ and $\chi_{\rm eff} \ga 0.3$ (orange region); and 3) mergers with $q \la 0.3$ and $\chi_{\rm eff} \la 0.5$ (green region). \textit{Right panel:} Mass ratio ($q=M_{2}/M_{1}$) as a function of $\chi_{\rm eff}$ for BBH mergers at the trap (red points) and the remaining bulk disk (blue points) in 100 realizations of the previous model, modified by: 1) suppressing mergers of binaries which form with $\vec{L}_{\rm b}$ anti-aligned with respect to $\vec{L}_{\rm d}$ (retrograde binaries, see also Figure \ref{fig:retrodiagram}); 2) decreasing the number of high mass BH relative to low mass BH, and modifying migration torques with a luminosity-dependent prescription from \citet{Hankla20}; and 3) segregating BH by mass such that more massive BH are more likely to be disk-embedded, and have their spin axis aligned with the disk (while less massive BH have lower spins that are initially isotropic). See text for details.

\label{fig:mergers}}
\end{figure*}

\citet{McK20a,McK20b} performed Monte-Carlo simulations of large numbers of BBH mergers in 1D models of AGN disks spanning a wide range of  assumptions.  Fig.~\ref{fig:mergers}(a) shows ($q,\chi_{\rm eff}$) for BBH mergers with assumptions from R7 in \citep{McK20a} with no discernible correlation in the simulated mergers. Here we revisit our assumptions to investigate what is required to generate a ($q,\chi_{\rm eff}$) anti-correlation for the AGN channel. It helps to guide our intuition to consider that an anti-correlation in ($q,\chi_{\rm eff}$) can appear only if we suppress mergers in specific parts of ($q,\chi_{\rm eff}$) parameter space. These regions, in turn, correspond to particular physical processes. We discuss three particular regions of parameter space (regions 1,2 \& 3 highlighted in Fig.~\ref{fig:mergers}(a)) below, where BBH mergers must be suppressed to obtain a correlation in $q$-$\chi_\mathrm{eff}$ similar to that observed by \citet{Tom2021}.

First, BBH mergers with negative $\chi_{\rm eff}$ (purple region of Fig.~\ref{fig:mergers}(a)) are suppressed with respect to positive $\chi_{\rm eff}$ mergers, except possibly around $q \sim 1$. In order to do this, we require binaries in AGN to  preferentially form either:1) biased towards both prograde spins \emph{and} prograde orbital angular momentum, or 2) biased towards both retrograde spins \emph{and} retrograde orbital angular momentum. Here we shall simply assume that `retrograde' binaries are ionized or softened at much higher rate than `prograde' binaries due to symmetry breaking (see below).
We do not include in our model orbiters moving  backwards through the disk gas (anti-parallel orbital angular momentum to $\vec{L}_{\rm d}$), since these orbits decay on short-timescales relative to the lifetime of an AGN disk \citep{Secunda20b}.

Second, BBH mergers with $q \ga 0.5$ but $\chi_{\rm eff} \ga 0.3$ (orange region of Fig.~\ref{fig:mergers} (a)) are suppressed. The simplest way of doing this in the AGN channel is to suppress the number of massive BH mergers relative to less massive BH mergers. This requires decreasing the relative number of more massive BH in the disk (e.g., steepening the slope of the BH mass function), or disrupting their mass-dependent migration, or both.

Third, mergers with $q \la 0.3$ and $\chi_{\rm eff} \la 0.3$ (green region in Fig.~\ref{fig:mergers}(a)) are suppressed. The simplest way of doing this in the AGN channel is to assume that spin orientations in encounters between more massive and less massive BH are not isotropically distributed. Rather, the more massive BH have spin orientations that prefer weak alignment with $L_{\rm disk}$, and consequently will be weakly aligned with the orbital angular momentum of binaries formed in the disk ($\vec{L}_{\rm b}$). This will tend to push $q \la 0.3$ mergers to higher positive $\chi_{\rm eff}$ values. We shall consider each of these suppressive effects in turn below.

\subsection{Softening or Ionizing retrograde binaries: suppressing $\chi_{\rm eff}<0$}
BBH mergers with negative $\chi_{\rm eff}$ imply that the projection of spins from masses $M_{1},M_{2}$ onto the binary orbital angular momentum about its center of mass ($\vec{L}_{\rm b}$) is net negative. i.e. the spin orientations are significantly anti-aligned with $\vec{L}_{\rm b}$. There is a relative deficit of mergers (in O3a) with $\chi_{\rm eff}<0$ \citep{o3a_pop}. In a gas-free dynamics merger channel, such as globular or nuclear clusters, an isotropic distribution of spin orientations should be expected and dynamical encounters should be spherically symmetric. So approximately half of the mergers in clusters would be expected to have negative $\chi_{\rm eff}$. In AGN disks, while binaries can form with $\vec{L}_{b}$ aligned or anti-aligned with the disk orbital angular momentum ($\vec{L}_{\rm d}$), the fate of the binary depends on the details and rates of gas or dynamical hardening. The preferential plane of the AGN gas disk provides a mechanism for breaking the symmetries of a gas-free `pure' dynamics channel, particularly at locations in disks with stalling orbits like migration traps.

Gas-hardening efficiency should decrease at modest binary separations. Binary hardening can then stall and the details of dynamical encounters with tertiary objects in AGN disks, including other binaries, becomes important \citep[e.g.][]{Leigh18,Samsing20,Tagawa20c}. Hard, stalled binaries in AGN migrate essentially as a single object with mass $M_{\rm bin}$, and encounter other migrators; only a few such close encounters are needed to merge the binary \citep{Leigh18}. Here we point out that there are a few places where symmetry of dynamical encounters may be broken which may contribute to a deficit of mergers with $\chi_{\rm eff}<0$. We caution that the details of the dynamical encounters will be important for our conclusions; 
there have not yet been sufficient large scale numerical experiments of encounters in the regime where orbital angular momentum, non-negligible energy of encounter $\Delta E$, and orbital element perturbation are considered together. 

First, and most obviously, the hard-soft boundary for retrograde and prograde binaries is different for encounters on a given 'side' of the binary. The relative velocity, $\Delta v$, for encounters with migrators arriving from the outer disk is larger for retrograde binaries (see  Fig.~\ref{fig:symmetrybreaking}). This effect is most obvious at migration traps in the inner disk where most encounters arrive from the outer disk. Thus, the symmetry of encounters between retrograde and prograde binaries may be broken at a migration trap where most encounters arrive from the outer disk.

Second, since more massive embedded objects migrate faster in disks, more massive objects `catch up' with less massive objects at smaller semi-major axis. More massive encounters with a binary will have a higher (lower) encounter velocity for retrograde (prograde) binaries. Conversely, if a binary `catches up' to less massive interior orbiters, the encounter velocity is lower (higher) for retrograde (prograde) binary orbits around its center of mass---see Figure \ref{fig:symmetrybreaking}. As a result, the symmetry breaking provided by the disk on the direction of typical disk encounter could bias hardening encounters to those between prograde binaries and other migrators. However, the degree of asymmetry of this effect depend on the details of the encounter. For example, if a migrating tertiary spends an equal amount of time on both sides of a binary then symmetry of interactions is restored. 

Third, 
the (complicated) details of gas torques may yield a difference in the hardening efficiency of retrograde versus prograde binaries \citep[e.g.][]{Baruteau11,Derdzinski21,Li21}. \citet{Baruteau11} find that retrograde binaries harden faster than prograde binaries; \citet{Li21} find that gas torques \textit{soften} prograde binaries and \textit{only} retrograde binaries will be preserved. If correct, the only way of preserving the bias against negative $\chi_{\rm eff}$ mergers in the AGN channel, would be if the BH in the disk have a bias towards negative spins. Interestingly, \citet{Adam21} suggest that stars embedded in AGN disks are driven to strongly negative spins, which might yield a population of BH forming in the disk with negative spins. However, we should be cautious about gas hardening results---in all cases, the hardening torques examined are operating on binaries near the hard-soft boundary, and while gas hardening may produce binaries with small semi-major axes, such binaries will likely stall before GW efficient merger. For stalled binaries, dynamical interactions must become involved to produce mergers, but could also ionize the binaries (see below). In addition, 
while \citet{Li21} do not produce hardened prograde binaries, they emphasize the critical effect of the detailed distribution of gas nearest to each of the binary partners in determining torque sign. Given 
given the effect of going from 2D to 3D simulations \citep[see e.g.][for work in the protoplanetary context]{2012ApJ...758L..42Z,2020arXiv200904345D}, and the lack of feedback in any binary hardening simulations to date (which will certainly change the distribution of gas near each BH), we should be cautious in taking present conclusions at face value.

One might also consider retrograde \textit{migrators} interacting with prograde binaries, but 
the gas disk provides a very strong symmetry breaking effect; as shown in \citet{Secunda20b}, single retrograde migrators will rapidly decay into the SMBH due to eccentricity pumping and will be unavailable to interact with any other disk components after a relatively short period $O(10^5)~$yr.

\subsection{Suppressing $q\ga 0.5$, $\chi_{\rm eff}\ga 0.3$ mergers}

In general, near equal mass, high $\chi_{\rm eff}$ mergers in the AGN channel come from the merger of hierarchical merger products, often at or near a migration trap. In order to suppress these mergers, we must decrease the relative number of high mass, high spin BH embedded in the disk (compared to the spherical distribution), or prevent these objects from efficiently finding each other, as generally happens if the migration trap is universal.
To reduce the number of of high-mass disk-embedded objects, we can give the disk embedded objects a flatter mass function than those in the spheroid population; to reduce the efficiency of trap mergers we can disrupt the smooth migration torques provided by the disk gas.
Migration disruption can occur if AGN disks are typically highly turbulent or if feedback from accretion onto the embedded population alters the migration torques. For example, \citet{Hankla20} showed that accretion onto orbiters embedded in a disk generates a heating torque that can act to drive the orbiters outwards, i.e., against the typical sense of Type I migration within disks; importantly, the strength of the counter-directional heating torque is related to the mass of the migrator through its accretion luminosity, which scales with mass.

\subsection{Spin alignment among heavier BH: suppressing $q\la0.3$, $\chi_{\rm eff} \la 0.3$}
Since the gas disk provides a preferential plane for binary formation, aligning the spin of the heavier BH with the orbital angular momentum of the binary ($\vec{L}_{\rm b}$) can be physically motivated by any mechanisms that preferentially align the spin of the primary with the orbital angular momentum of the disk ($\vec{L}_{\rm d}$, and hence the orbital angular momentum of the binary, $\vec{L}_{\rm b}$). 
A few percent ($\sim 1-10\%$) mass accretion is sufficient to torque the spin of a mis-aligned BH into alignment with a disk \citep{Bogdanovic07}. So any mechanism that leads to an \textit{angular} mass segregation, such that more massive BH spend more time in the gas disk (or form in the disk, or arrive there substantially before the less massive BH) will provide appropriate conditions for the embedded BH population (i.e. $\vec{\chi}_{1}$ is parallel to $\vec{L}_{\rm b}$).

Fortunately, if we begin with a spheroidal distribution of individual BH 
which has no preferred orbital plane and relatively small natal spins that are randomly aligned, the introduction of a gas disk will provide a torque that will align the orbits of BH with the gas disk. The dominant aligning torque is Bondi-Hoyle-Lyttleton drag, which depends on $M^2$, resulting in faster alignment of more massive BH orbits with the AGN disk \citep{Fabj20}. Such BH have  more time to accrete, and thus they have more time to have their (presumably random initial) spins aligned with the disk compared to their less massive partners. This mechanism requires the inner AGN disk to be dense enough that orbital capture is efficient within AGN lifetimes  \citep{Fabj20,MacLeodLin20}. Another possibility is that there is a preferential plane for AGN activity in galactic nuclei, such that angular mass segregation occurs due to leftover mergers from prior AGN episodes in the same plane \citep[e.g.][]{BertiVol08,Fan11}; or that mutual interactions even in the absence of a gas disk provide a preferential plane that results in disky assemblages of heavier BH at all times \citep[e.g.][]{Alexander07}, or that compact objects in the disk form from a top heavy IMF \citep[e.g.][]{Yuri03}, or are driven to higher masses through accretion \citep[][]{CantielloJermynLin2021}. Such mechanisms would enhance angular mass segregation and drive the preferential spin alignment of the heavier partners in BBH mergers.

\section{Modifications to Monte Carlo simulations}

Based on the phenomenological considerations above, we made some very simple modifications to the Monte Carlo simulations of \citet{McK20b} as described below. We emphasize again that these initial choices of parameters, while physically motivated, are only to illustrate that a population with a ($q,\chi_{\rm eff}$) anti-correlation can be produced by an AGN channel. Detailed parameter ranges will be studied in future work.

For initial mass distributions, we introduced 1) a disk mass function ($M^{-1}$) to correspond to a heavier disk BH population ([$20,50M_{\odot}$]). We also introduced 2) a spheroid BH mass function ($M^{-2}$) from which we randomly draw the lighter captured BH population ($[5,30]M_{\odot}$). We normalized both distributions so that $\sim 40$ BH from 1) start in the disk and $\sim 200$ BH from 2) are captured by the disk in $1 \rm{Myr}$, for each realization of the AGN disk.

For initial spin distributions, we assumed that the population in 2) had random spin orientation but a narrow range of dimensionless spin magnitudes parameterized by $a=s(1-s)$ where $s=[-0.98,0.98]$ is a random draw from a uniform distribution. This is not a physical model, but limits initial spin magnitudes to a draw between $a=[-0.25,0.25]$ among the captured BH population and corresponds to an assumption of small, but non-zero initial BH spin magnitude at birth. For the population in 1), we assumed that by starting off embedded within the AGN disk, BH had been slightly spun-up at $t=0$. We parameterized this by drawing initial spin magnitudes from a distribution given by $s(1-s) + 0.15$ where $s=[-0.98,0.98]$. Again, this is not a physical model, but has the effect of biasing initial BH dimensionless spin parameters in the disk to a draw between $a=[-0.1,0.35]$, corresponding to an assumption that BH have been marginally spun up, either from direct accretion from the disk, or possibly from previous AGN episodes in the same plane.
Our average spin bias of $s(1-s)+0.15$ selected here corresponds to around a Myr of gas accretion at the Eddington rate, or equivalently, shorter times at super-Eddington rates. The corresponding spin alignments in population 2) are random. 
A detailed study of the allowed range of spin-up by disks will be carried out in future work.

For accretion over the course of the simulation, we assumed that BH that are initially embedded in the disk accrete at slightly super-Eddington ($x2$) rates. This has the effect of driving a faster torquing of spin orientation into alignment with the AGN disk \citep{Bogdanovic07} over 1Myr, with a modest change in mass.

For migration, we assumed that feedback from accretion modifies the gas torques on embedded objects. We used the parameterization from \citep{Hankla20}, which approximates  the ratio of the heating torque ($\Gamma_{\rm heat}$) to the migration torque ($\Gamma_{\rm mig}$) is
\begin{equation}
    \frac{\Gamma_{\rm heat}}{\Gamma_{\rm mig}} \approx 0.07 \left( \frac{c}{v_{\rm orb}}\right) \epsilon \tau^{-1} \alpha^{-3/2}
\end{equation}
where $v_{\rm orb}$ is the orbital velocity of the migrator, $\epsilon$ is the Eddington ratio of accretion, $\tau$ is the disk optical depth and $\alpha$ is the disk viscosity parameter. The result is a net outward heating torque on embedded objects, which gets co-added to the (generally) inward migration torque and has the effect of slowing down the general drift inwards of population 1) BH. The heating torque depends strongly on the disk optical depth,  so a non-uniform quality now appears in the rate of migration of embedded objects, depending on their radial location in the disk.

For disk size, we assumed that the dense disk cuts off at $10^{4}r_{g}$ (or $0.05~$pc for $M_{SMBH}=10^8~M_{\odot}$). This has the effect of inhibiting the number of embedded objects from population 1) and therefore also reduces the relative number of mergers of within population 1). This size scale is larger than the O($10^{3}r_{g}$)inferred from reverberation mapping or microlensing \citep{Chartas16}, so could be revised in a parameter study, but is roughly where star formation is expected to consume the unstable outer disk \citep[e.g.][]{Sirko03}.

\begin{figure*}
\begin{center}
\includegraphics[width=0.75\linewidth,angle=0]{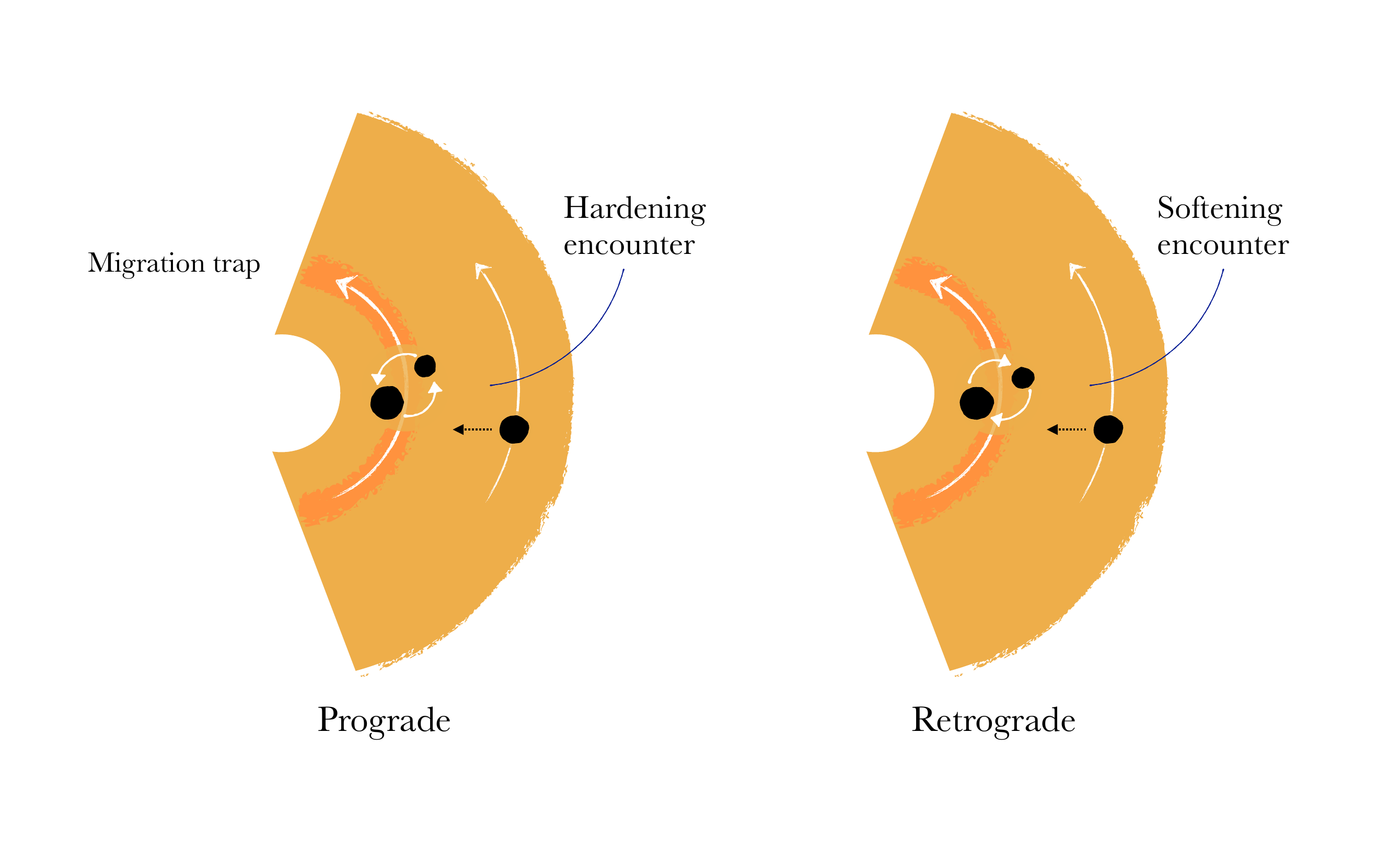}
\end{center}
\caption[BrokenSym]{
Breaking the symmetry of hardening/softening encounters: The hard-soft boundary is different for pro- and retro-grade binaries for encounters at larger and smaller disk radii. 
This symmetry breaking effect is strongest at migration traps (or other stalling orbits), since encounters with binaries at a trap will mostly be with migrators from larger disk radii.
\label{fig:symmetrybreaking}}
\end{figure*}

For dynamical interactions, we simply assumed that all binaries which \textit{would have} formed retrograde binaries, fail to form, and the individual components continue their migration undisturbed. We assumed that there is a migration trap in this disk model. The efficiency of the migration trap depends on the surface density change in the inner disk \citep{Bellovary16} as well as the effect of a Bondi accretion headwind \citep{Zhen21}. The latter depends on the details of gas flow around embedded objects, which remains poorly understood.  If an AGN disk exhbits large changes in surface density, particularly in the inner disk, where the pressure is high and disk mass is small, we should expect regions of the disk where in-migration can slow or stall and a pile-up can occur. Indeed in \citet{McK20b} we found that when we excluded the migration trap in a \citet{TQM05} disk model, following the treatment of \citet{Dittmann2020}, the change in disk surface density remained sufficient to substantially decrease migration torques. This allowed a large number of interactions and mergers in what we termed a migration `swamp'. Importantly, migration stalling strengthens the symmetry-breaking, binary ionization dynamic shown in Fig.~\ref{fig:symmetrybreaking}, since most encounters for a stalled binary are due to `catch-up' from larger disk radii.

Fig.~\ref{fig:mergers}b shows the results of implementing all of the above modifications in our simulations. The points in Fig.~\ref{fig:mergers}b correspond to all mergers in $100$ different realizations of a migration trap (red points) and bulk disk (blue points) in a \citet{Sirko03} model AGN disk around a $10^{8}M_{\odot}$ SMBH. The migration trap is located at $700 r_g$ and the disk has a lifetime of 1Myr. This most nearly matches run R7 in \citet{McK20b}, though the number of BH, their capture rate, their mass functions, and the existence of retrograde binaries have all been altered as discussed above.
A clear trend in ($q,\chi_{\rm eff}$) is apparent for the trap mergers (red points), although the overlap with the 1$\sigma$ distribution from \citep{Tom2021} is not particularly good. 
There is significant scatter about the trend, particularly at $q \ga 0.5$, but smaller scatter in $\chi_{\rm eff}$ at $q \la 0.5$ in trap mergers. We emphasize that our goal here is to establish what sort of phenomenological conditions are required in order to generate a possible ($q,\chi_{\rm eff}$) anti-correlation in Monte Carlo simulations of BBH mergers in AGN disks. While our initial assumptions here yield plausible results relatively similar to \citep{Tom2021}, a N-body study is required to test ionization dynamics and a detailed parameter space MC study is required to establish what conditions in AGN disks are ruled out by \citet{Tom2021}. In particular, since the trap mergers yield a stronger anti-correlation, it may be that we need to reduce the size of the bulk disk to $<10^{4}r_{g}$ to minimize the bulk disk contribution. Nevertheless, at a qualitative level, our results \textit{strongly} suggest that while gas torques may play an important role in forming BBH in AGN disks, dynamical interactions are necessary to reach a stage of GW-driven merger. This, in turn, suggests a modest stalling radius for gas-driven binary hardening torques, and the common existence of migration traps  in AGN disks.

\section{Discussion}
The possibility of an anti-correlation in ($q,\chi_{\rm eff}$) parameter space of BBH mergers is  exciting, since it will require distinct constraints on all of the proposed merger channels \citep{Tom2021}. Here we point out that the AGN channel has multiple conditions that may contribute to symmetry breaking in spherical mass distribution, binary formation and binary hardening in the disk.  

The key details underpinning a phenomenological anti-correlation between $(q,\chi_{\rm eff})$ in the AGN channel are symmetry-breaking in both the angular mass distribution of BH and in the hardening of binaries to merger (either dynamical or gas-driven). An AGN disk can be thought of as breaking the spherical symmetry of a gas-free galactic nucleus and promoting angularly dependent mass segregation. More massive BH are more quickly captured by the AGN disk \citep{Fabj20} and will therefore spend more overall time in the disk, spinning up and experiencing torquing towards alignment with the disk.
However, without the assumption of an offset for the spin magnitudes of the initial, disk-embedded, heavy BH population, we are unable to obtain such a clear anti-correlation among trap mergers. Thus, a separate possibility is that mass segregation in gas-free galactic nuclei leads to disky distributions of massive BH. Such disky distributions would provide a preferential plane for AGN disk formation, exchanging angular momentum with incoming gas. Or, a long-term preferential plane could be established if there are multiple AGN episodes delivered from the same fuel reservoir. Likewise, a disky distribution of BH implies the interval between fuelling episodes (duty cycle) is less than the relaxation timescale. The fuel reservoir (torus) can be volume filling and must therefore contain a fraction of the nuclear star cluster. Objects embedded in the torus will accrete, spin up and torque over time and would naturally end up embedded in inflowing gas. Yet another possibility is a top-heavy IMF for star formation in AGN disks which naturally promotes a more massive disky component \citep{Yuri03}. A combination of these mechanisms could account for a more massive disk component.

\begin{figure}
\begin{center}
\includegraphics[width=0.85\linewidth,angle=0]{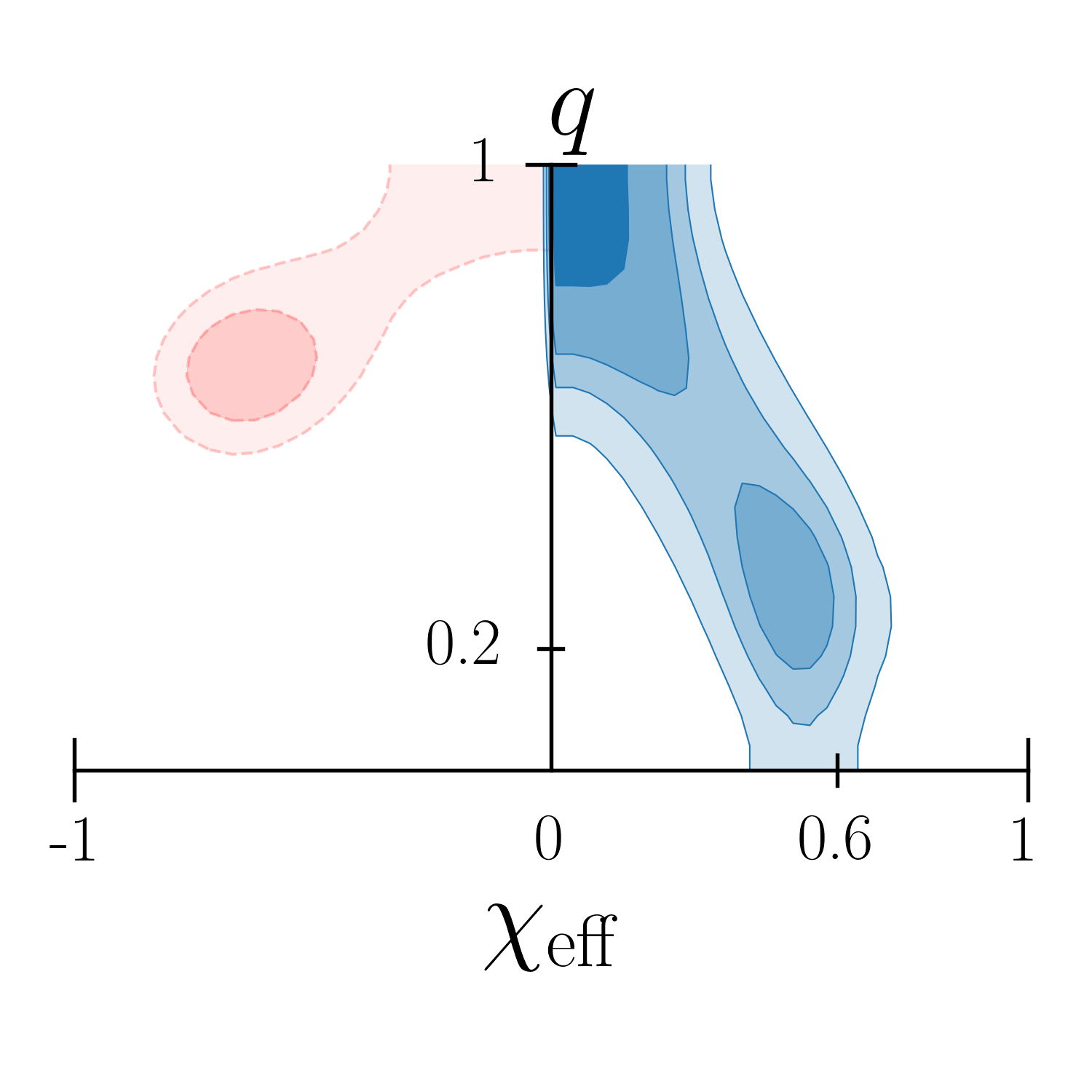}
\end{center}
\caption[predictions]{Cartoon representation of populations expected from this model. In blue are the $\chi_{\rm eff}>0$ populations as discussed in the text. In red are predictions for a (small) population of mergers due to formation of some BBH with $\chi_{\rm eff}<0$. If the hard-soft boundary symmetry breaking for retro- and pro-grade binaries is important, then we expect the slope of the $\chi_{\rm eff}<0$ merger population to be shallower than that for the $\chi_{\rm eff}>0$ population (see text). 
\label{fig:predictions}}
\end{figure}

Fig.~\ref{fig:predictions} is a cartoon summarizing (in blue) what we can naturally generate in AGN models, and (in red), \emph{predictions} for BBH mergers with negative $\chi_{\rm eff}$ from this channel. If retrograde binary ionization were irrelevant to BBH mergers in AGN disks then we should broadly expect a mirror image distribution of Fig.~\ref{fig:mergers}(b) at $\chi_{\rm eff}<0$ (i.e. mapping through $\chi_{\rm eff}=0$). However, if retrograde binary ionization is important for this channel, then there are two clear expectations: First, the absolute number of mergers with $\chi_{\rm eff}<0$ must be less than the corresponding number of mergers with $\chi_{\rm eff}>0$. Second, those retrograde binaries that survive to merger may be massive enough compared to dynamical encounters that they can survive. Thus, we should expect the slope of the ($q,\chi_{\rm eff}$) distribution in AGN to be shallower at $\chi_{\rm eff}<0$ than at $\chi_{\rm eff}>0$.  

We show that an anti-correlation in ($q,\chi_{\rm eff}$) parameter space can arise naturally in the AGN channel provided that a few phenomenological assumptions hold: 1) Heavier BH live in the AGN disk and tend to spin up into alignment with the disk. 2) Lighter BH live in the spheroid surrounding the AGN disk and have random spin alignments. 3) The inner AGN disk is dense, in order that BH from 2) are captured, but 4) not large radially, to limit the number of BH from population 1). 5) Migration must not be smooth, either due to a turbulent disk and/or feedback from the embedded objects, to cap the number of mergers between BH in 1). 6) Dynamical encounters within the disk must be common enough that binaries which orbit retrograde about their center of mass (anti-aligned with the disk angular momentum) are preferentially ionized. The ionization symmetry-breaking effect is strongest at stalled migration orbits, e.g. migration traps. Our understanding is that at present, no other BBH merger model channel can naturally produce such an anti-correlation. In particular, isolated binary evolution tends to lead to BH spins which are both aligned with the orbital angular momentum of the binary, and also tends towards $q\sim1$ mergers; similarly, globular cluster dynamics tends to sort the masses in binaries towards $q\sim1$ (through exchange interactions), and robustly produces symmetric distributions of $\chi_{\rm eff}$.

\section*{Acknowledgements.}
We would like to thank Imre Bartos for helpful feedback and comments. BM \& KESF are supported by NSF AST-1831415 and Simons Foundation Grant 533845.

\section*{Data Availability}
Any data used in this analysis are available on reasonable request from the first author (BM).

\bibliographystyle{mnras}
\bibliography{refs1} 

\end{document}